\documentclass[12pt,a4paper,twoside]{article}
\usepackage[utf8]{inputenc}
\usepackage[T2A]{fontenc}
\usepackage[english]{babel}
\usepackage{epsf,epsfig,rotating}
\usepackage{mathtools}
\usepackage{xcolor}
\usepackage{ulem}
\begin{document}

\begin{center}

{\Large \bf 
Radiative corrections to Higgs masses for the MSSM Higgs potential with the dimension-six operators
} \vspace{4 mm}

M.N. Dubinin$^*$, E.Yu. Petrova$^\dagger$

\vspace{4 mm}

{\it $^*$Skobeltsyn Institute of Nuclear Physics, Lomonosov Moscow State University,
119991 Leninskiye Gory, Moscow, Russian Federation}\\
{\it $^\dagger$Physics Department, Lomonosov Moscow State University,
119991 Leninskiye Gory, Moscow, Russian Federation}\\

\begin{abstract}
In the framework of the effective field theory approach to heavy supersymmetry radiative corrections in the Higgs sector of the Minimal Supersymmetric Standard Model (MSSM) for the effective potential decomposition up to the dimension-six operators are calculated.
Symbolic expressions for the threshold corrections induced by $F$- and $D$- soft supersymmetry breaking terms are derived and the Higgs boson mass spectrum respecting the condition $m_h=$125 GeV for the lightest $CP$-even scalar is evaluated. 
\end{abstract}

\end{center}

\vspace{4 mm}


\section{Introduction}

The absence of a signal of supersymmetric partners at the LHC up to the mass range of 1--2 TeV \cite{superpartners} increased an interest in the 'heavy supersymmetry' scenarios \cite{heavy_susy} of the Minimal Supersymmetric Standard Model (MSSM), where the condition $m_h=$125 GeV for the lightest $CP$-even scalar state, perhaps, observed by the ATLAS and CMS collaborations \cite{atlas_cms} is respected explicitly in the MSSM parameter space. Large radiative corrections to the MSSM two-Higgs doublet sector which raise up $m_h$ from the maximal tree-level value of $m_Z$ to the observable value of 125 GeV appear due to large values of soft supersymmetry breaking parameters, which are associated with large masses of third generation quark supersymmetric partners, large mixing of supersymmetric partners, and restricted from above by the availability of perturbative regime. For this reason acceptable domains of the MSSM parameter space are rather limited \cite{limited} although there are several variants of such 'fine tuning'. In order to ease tensions of parametric scenarios of the MSSM, two ways of action are appropriate, first, more precise calculations of radiative corrections at higher 
loops/decomposition
of the effective potential in the higher inverse powers of $M_{S}$ (i.e. including effective operators $1/M^n_{S}\; O(\Phi^{n+4})$ in the decomposition of Coleman-Weinberg type potential); second, the transition to extensions of the MSSM. For example, extensions of the MSSM where the superpotential includes an additional chiral singlet field (Next to Minimal Supersymmetric Standard Model (NMSSM) \cite{nmssm}), or more chiral fields, are known. It is assumed that some new physics beyond the MSSM exists at an energy scale which is not too far away. Probably, such scale of the order of 10$^1$ TeV is somewhat higher than the mass scale of superpartners $M_{S}$. 

In the framework of a picture where MSSM is a low-energy limit of an extended theory (not only NMSSM, for example, supersymmetric grand unification models \cite{grand} or supersymmetric left-right models \cite{left-right}) all possible effective operators of higher dimension should be introduced with the following separation of the observables which are sensitive to effects of the extended theory for phenomenological analysis. The effective Lagrangian of the MSSM extension can be written as a sum of operators suppressed by inverse powers of the new physics scale $M^{-1}$ and $M^{-2}$, each of which is $SU(3)_c \times SU(2)_L \times U(1)_Y $ invariant and respects $R$-parity. In the extended theory such operators are generated either at the tree-level or at the loop-level. It was expected that contributions of the tree-level operators to the specific observables were more important because the loop-level operators have an additional suppression factors proportional to $1/16\pi^2$. However, additional enhancements by large MSSM parameters (like $\tan \beta = v_2/v_1$ which can compensate also an extra power of the mass scale $M$)  make the situation with various contributions rather nontrivial. A number of studies prior to Higgs boson discovery can be found in the literature. A complete list of the tree-level dimension-five and dimension-six effective operators can be found in \cite{piriz}. Note that supersymmetry restricts possible effective operator categories, for example, no operators of dimension-five involving Higgs-Higgsino supermultiplet and gauge-gaugino supermultiplet exist since no gauge invariant form can be constructed using three MSSM chiral superfields. Analogously, no operators of dimension-six involving Higgs-Higgsino supermultiplet exist because operators of this type must contain five chiral superfields, apparently, such forms violate gauge invariance. Various aspects related to an extensions by the dimension-five/dimension-six operators were systematically analyzed in \cite{dimension_five_six}. 

As mentioned above, radiative corrections coming from the loop diagrams with top quark and top superpartner are very important \cite{mssm_radcor_1, mssm_radcor_2} both for large $\tan \beta$ and small $\tan \beta$ parameter. The tree-level mass of the lightest $CP$-even state $h$ is maximized at large $\tan \beta$. For small trilinear parameters $A_{t,b}$ and large stop mass scale $M_{S}$ when $A_t/M_{S}$ and $\mu/M_{S}$ are less than one (in other words, in the case of moderate stop mixing parameter $X_t=A_t-\mu \cot \beta$), the correction to $m_h$ at the one-loop is controlled by the logarithm $\log M_{S}/m_{top}$ which is large enough for $M_{S}$ of the order of 10--100 TeV. For large trilinear parameters $A_{t,b}$ (or in the case of large stop mixing parameter) the correction is maximized at $A_t=M_{S} \sqrt{6}$ (so-called 'maximal mixing scenario' at the one-loop), and much smaller $M_{S}$ values of the order of 1 TeV are appropriate. At $\tan \beta \sim 1$ or even smaller, large mixing may appear due to large Higgs superfield mass parameter $\mu$ of about 10 TeV. Nontrivial interplay of $A_{t,b}, \mu, M_{S}$ and $\tan \beta$ parameters at the level of one-loop resummed Higgs potential was analyzed in detail for the potential decomposition in the inverse powers of $M_{S}$ up to operators of the dimension-four. The case of small mass splittings for quark superpartners \cite{carena95} was generalized for the situation when each stop and sbottom is independently decoupled at its specific mass scale \cite{mssm_radcor_new}
for some special MSSM effective potentials. Note that the two-loop effects may be included by using a renormalization group improvement of the effective potential \cite{primary_improvement}. The scale dependence of the one-loop result is reduced if the two-loop renormalization group improvement of the one-loop effective potential is accounted for \cite{carena95, improvement}.

In this paper the effective MSSM Higgs potential decomposition up to operators of the dimension-six involving scalars only is considered. Contribution of the dimension-six operators to observables can be separated insofar, as already mentioned above, the dimension-six operators involving only scalar isodoublets appear at the loop-level only. In Section 2 the mass basis for extended Higgs sector is constructed.
In Section 3 analytical expressions for the threshold corrections are derived and
some numerical evaluations for the mass spectra are performed.

\section{Mass basis for the case of effective potential with the dimension-six terms}

In this section we construct the basis for the mass states of physical scalars following \cite{own1}, where the case of dimension-four operators has been considered. Two Higgs doublets of the form 
\begin{eqnarray}
\label{dublets} \Phi_i= \left(\begin{array}{c} \phi_i^+(x) \\ \phi_i^0(x) \end{array} \right)=\left(\begin{array}{c} -i \omega_i^+ \\ \frac{1}{\sqrt{2}} (v_i+\eta_i+i \chi_i) \end{array} \right), \qquad
i=1,2
\end{eqnarray}
are used to define the general two-Higgs doublet potential which includes the dimension-six terms. The potential can be written as
\begin{equation}
\label{U}
U=U^{(2)}+U^{(4)}+U^{(6)},
\end{equation}
where the upper index shows the operator dimension in fields,
\begin{eqnarray}
U^{(2)} &=&
- \, \mu_1^2 (\Phi_1^\dagger\Phi_1) - \, \mu_2^2 (\Phi_2^\dagger
\Phi_2) - [ \mu_{12}^2 (\Phi_1^\dagger \Phi_2) +h.c.], \\
\label{U4}
U^{(4)} &=& \lambda_1
(\Phi_1^\dagger \Phi_1)^2
      +\lambda_2 (\Phi_2^\dagger \Phi_2)^2
+ \lambda_3 (\Phi_1^\dagger \Phi_1)(\Phi_2^\dagger \Phi_2) +
\lambda_4 (\Phi_1^\dagger \Phi_2)(\Phi_2^\dagger \Phi_1)+\\
&+& [\lambda_5/2
       (\Phi_1^\dagger \Phi_2)(\Phi_1^\dagger\Phi_2)+ \lambda_6
(\Phi^\dagger_1 \Phi_1)(\Phi^\dagger_1 \Phi_2)+\lambda_7 (\Phi^\dagger_2 \Phi_2)(\Phi^\dagger_1 \Phi_2)+h.c.], \nonumber \\
\label{U6}
U^{(6)} &=& \kappa_1 (\Phi^\dagger_1 \Phi_1)^3
+\kappa_2 (\Phi^\dagger_2 \Phi_2)^3+
\kappa_3 (\Phi^\dagger_1 \Phi_1)^2 (\Phi^\dagger_2 \Phi_2)+\kappa_4 (\Phi^\dagger_1 \Phi_1) (\Phi^\dagger_2 \Phi_2)^2+\\
&+&\kappa_5 (\Phi^\dagger_1 \Phi_1) (\Phi^\dagger_1 \Phi_2) (\Phi^\dagger_2 \Phi_1)+
\kappa_6 (\Phi^\dagger_1 \Phi_2) (\Phi^\dagger_2 \Phi_1) (\Phi^\dagger_2 \Phi_2)+ \nonumber \\
&+& [\kappa_7 (\Phi^\dagger_1 \Phi_2)^3+\kappa_8 (\Phi^\dagger_1 \Phi_1)^2 (\Phi^\dagger_1 \Phi_2)+\kappa_9 (\Phi^\dagger_1 \Phi_1) (\Phi^\dagger_1 \Phi_2)^2+ \nonumber \\
&+&\kappa_{10} (\Phi^\dagger_1 \Phi_2)^2 (\Phi^\dagger_2 \Phi_2)+
\kappa_{11} (\Phi^\dagger_1 \Phi_2)^2 (\Phi^\dagger_2 \Phi_1)+
\kappa_{12} (\Phi^\dagger_1 \Phi_2) (\Phi^\dagger_2 \Phi_2)^2+\nonumber \\
&+&\kappa_{13} (\Phi^\dagger_1 \Phi_1) (\Phi^\dagger_1 \Phi_2) (\Phi^\dagger_2 \Phi_2)+h.c.], \nonumber
\end{eqnarray}
so the parameters $\mu_1,\mu_2$ and $\mu_{12}$ are of dimension one, $\lambda_i, i=1,...7$ are dimensionless and the dimension of $\kappa_i, i=1,...13$ is of inverse mass squared. In the general case $\mu_{1}^2, \mu_2^2$, $\lambda_1,...,\lambda_4$ and $\kappa_1,...,\kappa_6$ are real, all other parameters can be complex. In this Section the mass basis for the general case of explicitly $CP$-violating potential \cite{own1, explicit_cp} with nonzero imaginary parts of $\mu$, $\lambda$ and $\kappa$ parameters will be constructed. Transformations of the $SU(2)$ states $\eta_{1,2}, \chi_{1,2}, \omega^\pm_{1,2}$, Eq.~(1), to the mass states $h$, $H$, $A$, $H^\pm$, $G^0$, $G^\pm$ can be performed using two orthogonal rotations
\begin{eqnarray}
\left( \begin{array}{c} \eta_1 \\ \eta_2 \end{array} \right)={\cal O}_\alpha
\left( \begin{array}{c} h \\ H  \end{array} \right), \qquad
\left( \begin{array}{c} \chi_1 \\ \chi_2 \end{array} \right)={\cal O}_\beta
\left( \begin{array}{c} G^0 \\ A  \end{array} \right), \qquad
\left( \begin{array}{c} \omega_1^\pm \\ \omega_2^\pm \end{array} \right)={\cal O}_\beta
\left( \begin{array}{c} G^\pm \\ H^\pm  \end{array} \right),
\end{eqnarray}
where
\begin{equation}
{\cal O}_X=\left( \begin{array}{cc} \cos X & -\sin X \\ \sin X & \cos X  \end{array} \right), \qquad
X=\alpha, \beta,
\end{equation}
(in the following we denote $\cos X=c_X$, $\sin X=s_X$, etc.) and the Higgs potential (\ref{U}) takes the form
\begin{equation}
\label{Umass}
U=c_0 A+c_1 h A+c_2 H A+\frac{m_h^2}{2} h^2+\frac{m_H^2}{2}H^2+\frac{m_A^2}{2} A^2+m_{H^\pm}^2 H^+H^-+I_3+I_4+I_5+I_6.
\end{equation}
Here $I_{3,4,5,6}$ denote the interaction terms of physical scalars and the coefficients $c_i$, $i=$0,1,2, which are dependent on the imaginary parts of $\lambda_i$ and $\kappa_i$ 
\begin{eqnarray}
c_1 &=& v^2(-1/2 \cdot {\rm Im}\lambda_5 c_{\alpha+\beta}+{\rm Im}\lambda_6 s_\alpha c_\beta-{\rm Im}\lambda_7 c_\alpha s_\beta)\\ \nonumber
&+&\frac{v^4}{4} (-c_{\alpha+\beta} s_{2 \beta} (3 {\rm Im} \kappa_7+{\rm Im} \kappa_{11}+{\rm Im} \kappa_{13})
+4(s_\alpha c_\beta^3 {\rm Im}\kappa_8-c_\alpha s_\beta^3 {\rm Im} \kappa_{12})\\ \nonumber
&+&2( s_\beta^2 (-3 c_\alpha c_\beta+s_\alpha s_\beta){\rm Im}\kappa_{10}
-c_\beta^2 (c_\alpha c_\beta-3 s_\alpha s_\beta){\rm Im}\kappa_{9})),
\\
c_2 &=& -\frac{v^2}{2} \{ {\rm Im}\lambda_5 s_{\alpha+\beta}+2 ({\rm Im} \lambda_6 c_\beta c_\alpha+{\rm Im}\lambda_7 s_\beta s_\alpha)\\
&+& v^2[2{\rm Im}\kappa_8 c_\beta^3 c_\alpha+{\rm Im}\kappa_9 c_\beta^2 (s_{\alpha+\beta}+2c_\alpha s_\beta)+{\rm Im}\kappa_{10} s_\beta^2 (s_{\alpha+\beta}+2c_\beta s_\alpha)\nonumber \\
&+& 2 {\rm Im}\kappa_{12} s_\beta^3 s_\alpha +\frac{1}{2}(3{\rm Im}\kappa_{7}+{\rm Im}\kappa_{11}+{\rm Im}\kappa_{13})s_{2 \beta} s_{\alpha+\beta}] \} \nonumber
\end{eqnarray} 
are equal to zero in the mass basis.
In a local minimum where derivatives of the potential in the fields are zero, $\mu_{1}^2$ and $\mu^2_2$ can be expressed as
\begin{eqnarray}
\label{mu1}
\mu_1^2 &=& -{\rm Re} \mu_{12}^2 t_\beta+\frac{v^2}{4} (4 \lambda_1 c_\beta^2 +3 {\rm Re}\lambda_6 s_{2 \beta} +2 s_\beta^2 (\lambda_{345}+{\rm Re} \lambda_7 t_\beta ))+\\
&+&\frac{v^4}{4}(3 \kappa_1 c_\beta^4+5{\rm Re}\kappa_8 c_\beta^3 s_\beta+3({\rm Re}\kappa_7+{\rm Re}\kappa_{11}+{\rm Re}\kappa_{13})c_\beta s_\beta^3+\nonumber \\
&+& ({\rm Re} \kappa_9+(\kappa_3+\kappa_5)/2)s_{2 \beta}^2+(\kappa_4+\kappa_6+2 {\rm Re}\kappa_{10}+{\rm Re}\kappa_{12} t_\beta)s_\beta^4), \nonumber \\
\label{mu2}
\mu_2^2 &=& -{\rm Re} \mu_{12}^2 \cot \beta+\frac{v^2}{4} (4 \lambda_2 s_\beta^2 +3 {\rm Re}\lambda_7 s_{2 \beta} +2 c_\beta^2 (\lambda_{345}+{\rm Re} \lambda_6 \cot_\beta ))+\\
&+&\frac{v^4}{4}(3 \kappa_2 s_\beta^4+5{\rm Re}\kappa_{12} s_\beta^3 c_\beta+3({\rm Re} \kappa_7+{\rm Re}\kappa_{11}+{\rm Re}\kappa_{13})s_\beta c_\beta^3+\nonumber \\
&+& ({\rm Re} \kappa_{10}+(\kappa_4+\kappa_6)/2)s_{2 \beta}^2+(\kappa_3+\kappa_5+2 {\rm Re}\kappa_{9}+{\rm Re}\kappa_{8} \cot_\beta)c_\beta^4). \nonumber 
\end{eqnarray}
The real part of $\mu^2_{12}$ is fixed by zero eigenvalue of the mass matrix (which ensures massless Goldstone boson state and defines the $CP$-odd scalar mass $m^2_A$)
\begin{eqnarray}
\label{re12}
{\rm Re} \mu_{12}^2&=&s_\beta c_\beta \left(m_A^2+\frac{v^2}{2} (2 {\tt Re}\lambda_5+{\tt Re} \lambda_6 \cot\beta +{\tt Re}\lambda_7 \tan\beta)\right)+\\
&+& v^4 \left( {\rm Re}\kappa_9 c_\beta^3 s_\beta+{\rm Re}\kappa_{10} c_\beta s_\beta^3 +\frac{1}{4}[{\rm Re}\kappa_8 c_\beta^4+{\rm Re} \kappa_{12} s_\beta^4+(9{\rm Re}\kappa_7+{\rm Re}\kappa_{11}+{\rm Re}\kappa_{13}) s_\beta^2 c_\beta^2] \right) \nonumber
\end{eqnarray}
The requirement $c_0=0$ in Eq.~(\ref{Umass}) fixes the imaginary part of $\mu^2_{12}$
\begin{eqnarray}
\label{im12}
{\rm Im}\mu_{12}^2 &=& \frac{v^2}{2} (s_\beta c_\beta {\rm Im} \lambda_5+c_\beta^2 {\rm Im}\lambda_6+s_\beta^2 {\rm Im}\lambda_7)+\\
&+& \frac{v^4}{4} ({\rm Im} \kappa_8 c_\beta^4+2 {\rm Im}\kappa_9 c_\beta^3 s_\beta+(3 {\rm Im}\kappa_7+{\rm Im}\kappa_{11}+{\rm Im}\kappa_{13}) c_\beta^2 s_\beta^2+2 {\rm Im}\kappa_{10} c_\beta s_\beta^3+{\rm Im}\kappa_{12} s_\beta^4) \nonumber
\end{eqnarray}
Minimization conditions above must be performed for a generic two-doublet potential. In the following the case of MSSM potential will be analyzed. The one-loop resummed MSSM potential at the renormalization scale $m_{top}$ using dimensional reduction and the $\overline{MS}$-scheme can be written in the form
\begin{equation}
\label{eff}
U_{\rm eff}=U^0+\frac{3}{32 \pi^2}{\rm tr} {\cal M}^4 \left( \ln \frac{{\cal M}^2}{m_{top}^2}-\frac{3}{2} \right),
\end{equation}
where $U^0$ is a tree-level potential at the scale $M_{S}$ 
\begin{eqnarray}
U^0&=&- \, \mu_1^2 (\Phi_1^\dagger\Phi_1) - \, \mu_2^2 (\Phi_2^\dagger
\Phi_2) - [ \mu_{12}^2 (\Phi_1^\dagger \Phi_2) +h.c.]\\
&+&\frac{g_1^2+g_2^2}{8}[(\Phi_1^\dagger \Phi_1)^2+(\Phi_2^\dagger \Phi_2)^2]+\frac{g_2^2-g_1^2}{4} (\Phi_1^\dagger \Phi_1)(\Phi_2^\dagger \Phi_2)
-\frac{g_2^2}{4}(\Phi_1^\dagger \Phi_2)(\Phi_2^\dagger \Phi_1), \nonumber
\end{eqnarray}
and ${\cal M}^2={\cal M}^2_M+{\cal M}^2_\Gamma+{\cal M}^2_\Lambda$ is the squark mass matrix squared (see the Appendix). At the mass scale of quark superpartners the mass matrix elements are
\begin{equation}
\label{mijtree}
{\cal M}_{11}^2=m_A^2 s_\beta^2+m_Z^2 c_\beta^2, \qquad
{\cal M}_{22}^2=m_A^2 c_\beta^2+m_Z^2 s_\beta^2, \qquad
{\cal M}_{12}^2=-s_\beta c_\beta(m_A^2+m_Z^2).
\end{equation}
Radiative corrections to these tree-level expressions are parametrized using
\begin{eqnarray}
\label{lthrs}
\lambda_i(M)&=&\lambda_i^{\tt tree}({M_{S}})-\Delta \lambda_i(M)/2, \quad i=1,2,\\ 
\lambda_i(M)&=&\lambda_i^{\tt tree}({M_{S}})-\Delta \lambda_i(M), \quad i=3,...7, \nonumber 
\end{eqnarray}
\begin{center}
$\lambda_{1,2}^{\tt tree}=\frac{g_1^2+g_2^2}{8}, 
\lambda_{3}^{\tt tree}=\frac{g_2^2-g_1^2}{4},
\lambda_{4}^{\tt tree}=-\frac{g_2^2}{2},
\lambda_{5,6,7}^{\tt tree}=0,
\kappa_{i}^{\tt tree}=0,$ $i=1,..13$,
\end{center}
so corrections to the matrix elements of $CP$-even states mass matrix are
\begin{eqnarray}
\label{mijrad}
\Delta{\cal M}_{11}^2&=&-v^2 (\Delta \lambda_1 c_\beta^2+{\tt Re} \Delta \lambda_5 s_\beta^2+{\tt Re} \Delta \lambda_6 s_{2 \beta})+\\
&+& v^4 [3 \kappa_1 c_\beta^4+4 {\rm Re}\kappa_8 c_\beta^3 s_\beta+(\kappa_3+\kappa_5+3{\rm Re}\kappa_9)c_\beta^2 s_\beta^2 +\nonumber  \\
&+& (3 {\rm Re}\kappa_7+{\rm Re}\kappa_{11}+{\rm Re}\kappa_{13})c_\beta s_\beta^3+{\rm Re}\kappa_{10}s_\beta^4], \nonumber \\
\Delta{\cal M}_{22}^2&=&-v^2 (\Delta \lambda_2 s_\beta^2+{\tt Re} \Delta \lambda_5 c_\beta^2+{\tt Re} \Delta \lambda_7 s_{2 \beta})+  \\ 
&+& v^4[ {\rm Re}\kappa_9 c_\beta^4+(3 {\rm Re}\kappa_7+{\rm Re}\kappa_{11}+{\rm Re}\kappa_{13})c_\beta^3 s_\beta + \nonumber\\
&+& (\kappa_4+\kappa_6+3 {\rm Re}\kappa_{10})c_\beta^2 s_\beta^2 +4 {\rm Re}\kappa_{12}c_\beta s_\beta^3+3\kappa_2 s_\beta^4], \nonumber \\
\Delta{\cal M}_{12}^2&=&-v^2 (\Delta \lambda_{34} s_\beta c_\beta +{\tt Re} \Delta \lambda_6 c_\beta^2+{\tt Re} \Delta \lambda_7 s_\beta^2) + \\
&+&v^4 [{\rm Re}\kappa_8 c_\beta^4+(\kappa_3+\kappa_5+{\rm Re}\kappa_9)c_\beta^3 s_\beta+ \nonumber \\
&+&2({\rm Re} \kappa_{11}+{\rm Re} \kappa_{13})c_\beta^2 s_\beta^2+(\kappa_4+\kappa_6+{\rm Re}\kappa_{10})c_\beta s_\beta^3+{\rm Re}\kappa_{12}s_\beta^4].  \nonumber   
\end{eqnarray}
Then the masses of $CP$-even scalars can be expressed as
\begin{eqnarray}
\label{mHh}
m_{H,h}^2 = \frac{1}{2} (m_A^2+m_Z^2+\Delta{\cal M}_{11}^2+\Delta{\cal M}_{22}^2 \pm \sqrt{m_A^4+m_Z^4-2 m_A^2 m_Z^2 c_{4 \beta}+C}), 
\end{eqnarray}
where
\begin{equation}
\label{C}
C=4 \Delta{\cal M}_{12}^4+(\Delta{\cal M}_{11}^2-\Delta{\cal M}_{22}^2)^2-2 (m_A^2-m_Z^2)(\Delta{\cal M}_{11}^2-\Delta{\cal M}_{22}^2) c_{2 \beta}-4(m_A^2+m_Z^2) \Delta{\cal M}_{12}^2 s_{2 \beta},
\end{equation}
and the mixing angle $\alpha$ is defined by
\begin{equation}
\label{tg}
\tan 2 \alpha = \frac{2 \Delta {\cal M}_{12}^2-(m_Z^2+m_A^2)s_{2 \beta}}{(m_Z^2-m_A^2) c_{2 \beta}+\Delta {\cal M}_{11}^2-\Delta{\cal M}_{22}^2}.
\end{equation}
The $CP$-odd scalar mass $m_A$ can be expressed through $m_h$. Using Eq. (\ref{mHh}) one can define $m_A$ as an internal model parameter if the numerical value of the Higgs mass $m_h=$125 GeV is fixed 
\begin{equation}
m^2_A=\frac{m^2_h (C_1-m^2_h)+m^2_Z(C_2-C_3)-\Delta{\cal M}^2_{11} \Delta{\cal M}^2_{22} + \Delta{\cal M}^4_{12}}
             {C_1-C_2-C_3+m^2_Z c^2_{2\beta}},
\label{m2A}
\end{equation}
where
\begin{eqnarray*}
C_1&=&\Delta{\cal M}^2_{11}+\Delta{\cal M}^2_{22},\\
C_2&=&m^2_h-\Delta{\cal M}^2_{12}s_{2\beta},\\ 
C_3&=&\Delta{\cal M}^2_{11} s^2_\beta+\Delta{\cal M}^2_{22} c^2_\beta .
\end{eqnarray*}
The mass of charged Higgs boson in the form
\begin{eqnarray}
m_{H^\pm}^2&=&m_W^2+m_A^2-\frac{v^2}{2} ({\rm Re}\Delta \lambda_5-\Delta \lambda_4)+\\
&+& \frac{v^4}{4} [c_\beta^2(2 {\rm Re}\kappa_9-\kappa_5)+s_\beta^2(2 {\rm Re}\kappa_{10}-\kappa_6)-s_{2 \beta} ({\rm Re}\kappa_{11}-3{\rm Re}\kappa_7)] \nonumber
\end{eqnarray}
can be obtained diagonalizing the
corresponding mass matrix. Two important conditions which restrict implicitly the MSSM parameter space follow from Eq.~(\ref{mHh}):
\begin{equation}
\label{2cond}
m^4_A+m^4_Z-2m^2_A m^2_Z c_{4\beta}+C \geq 0; \quad
m^4_A+m^4_Z+\Delta{\cal M}^2_{11}+\Delta{\cal M}^2_{22}-2m^2_h \geq 0.
\end{equation}
\begin{figure*}[h]
\centering
\begin{minipage}[h]{0.45\linewidth}
\center{\includegraphics[width=\linewidth]{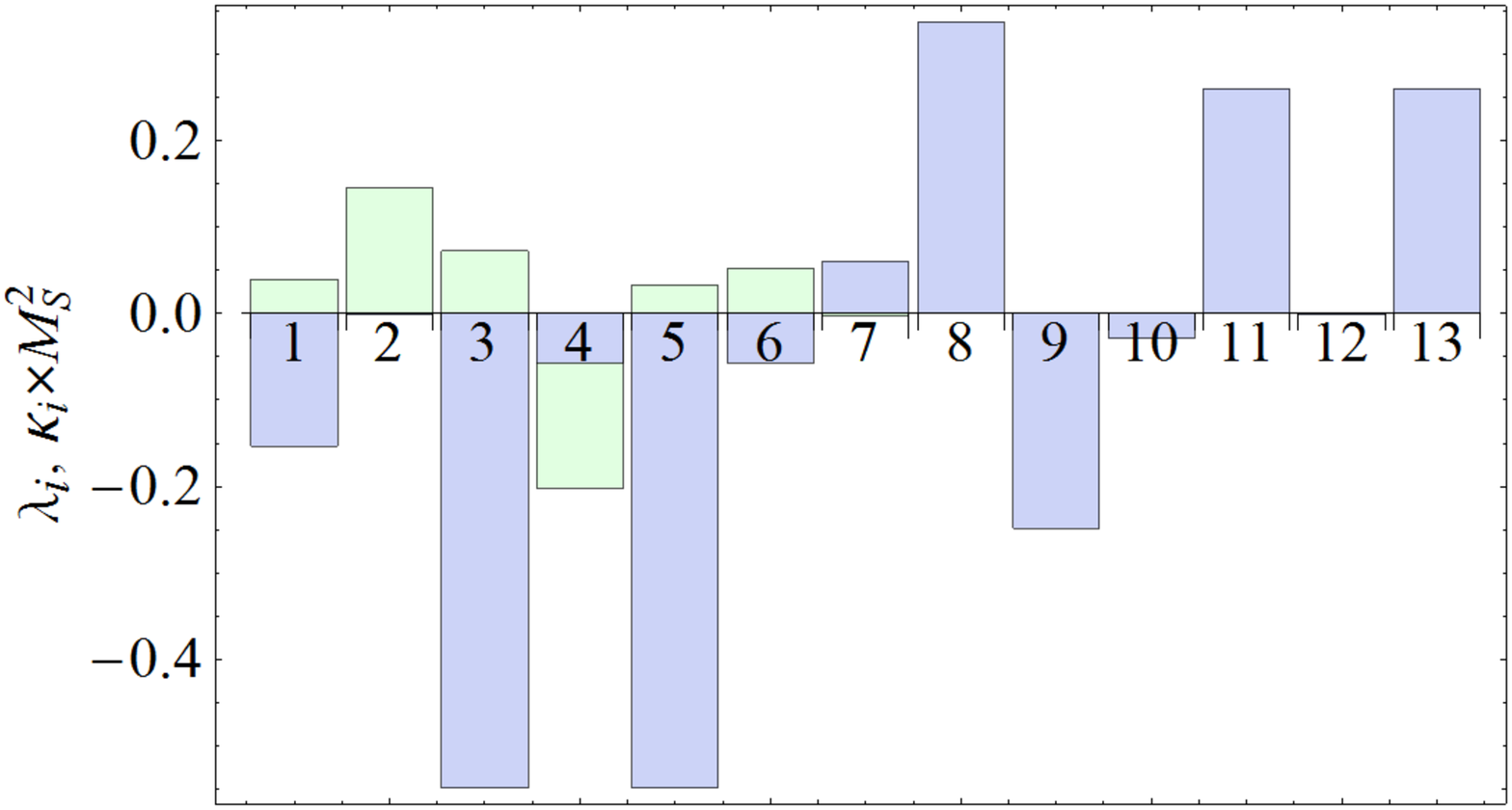}} (a)
\end{minipage} 
\begin{minipage}[h]{0.45\linewidth}
\center{\includegraphics[width=\linewidth]{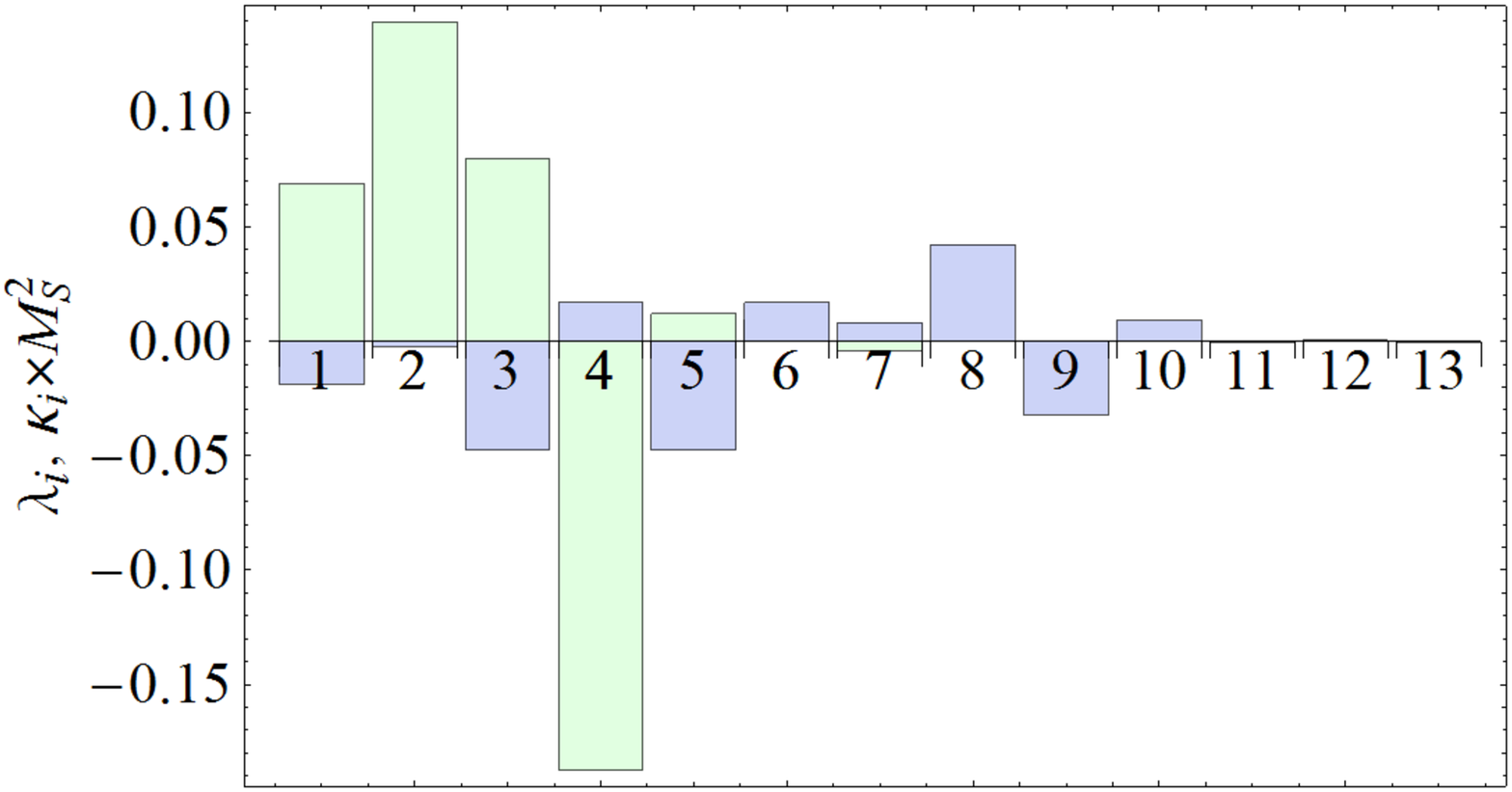}} (b)
\end{minipage} 
\caption{
The dimensionless parameters $\lambda_i$, $i=1,...7$ (green) calculated using the analytical results of \cite{own1} and
$\kappa_j \cdot M_S^2$, $j=1,...13$ (blue) calculated at the squark mass scale $M_S=5$ TeV (a) and $M_S=7$ TeV (b) for $A_t=A_b=10$ TeV, $\mu=14$ TeV, $\tan \beta=5$ .
}
\label{fig-1} 
\end{figure*}

\section{Symbolic expressions for $\kappa_i$ and numerical results}

\begin{figure*}[h]
\begin{minipage}[h]{0.5\linewidth}
\center{\includegraphics[width=1\linewidth]{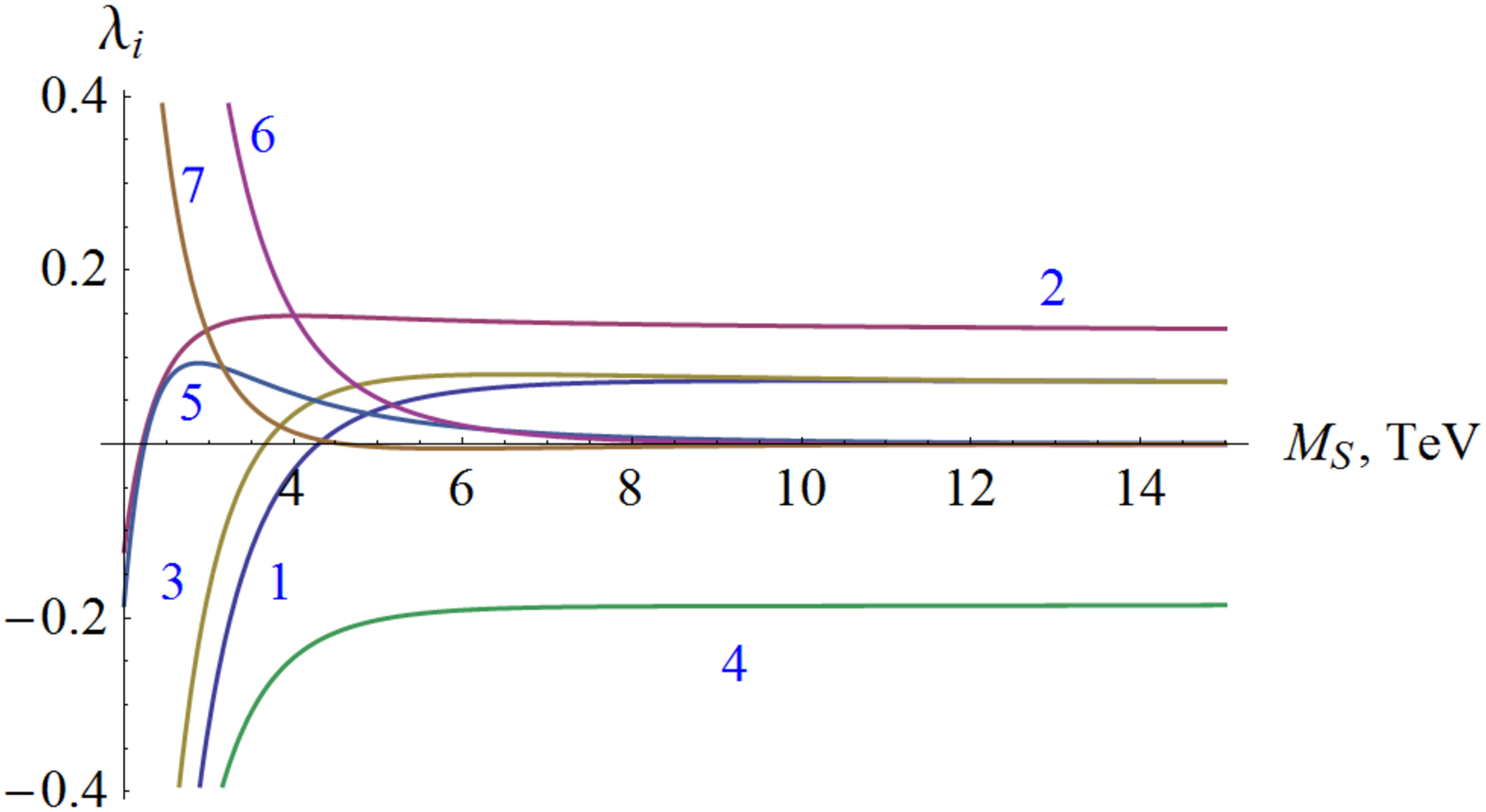}} (a)
\end{minipage} 
\begin{minipage}[h]{0.5\linewidth}
\center{\includegraphics[width=1\linewidth]{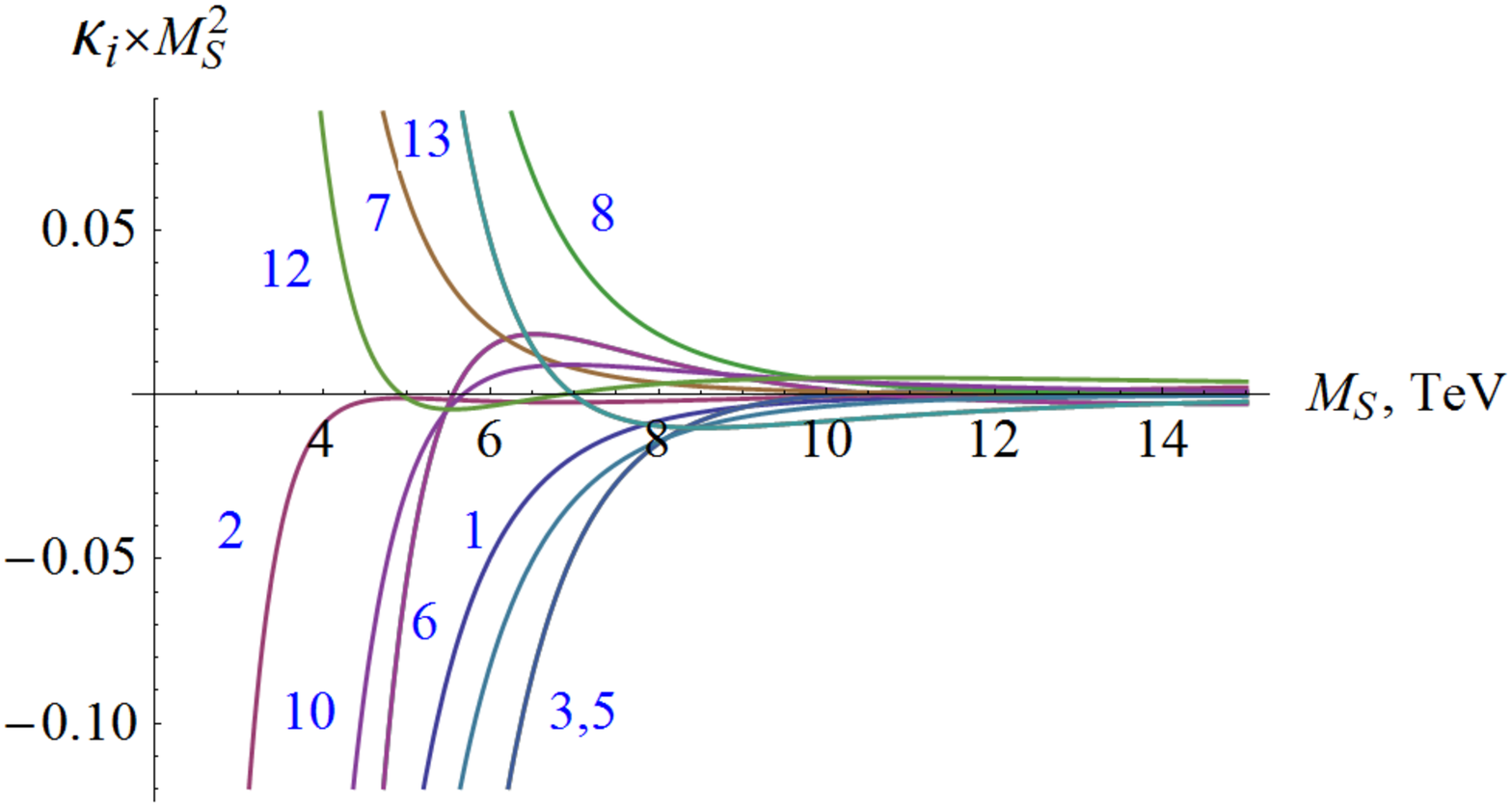}} (b)
\end{minipage}
\caption{
The dimensionless parameters $\lambda_i$ (a) and $\kappa_i \cdot M_S^2$ (b) as a function of $M_S$ for $A_t=A_b=10$ TeV, $\mu=14$ TeV, $\tan \beta=5$. $\lambda_i$ are evaluated using analytical formulae from \cite{own1}, where the contribution of nonleading $D$-terms is accounted for.
}
\label{fig-2} 
\end{figure*}

The one-loop expressions for parameters $\kappa_i$ in front of the dimension-six terms can be obtained decomposing the effective potential (\ref{eff}) in the inverse powers of $M_{S}$ in the approximation of degenerate squark masses \cite{carena95, haberhempfling}. In the following we are using the notation $M_{S}=M_{\tilde{Q}, \tilde{U}, \tilde{D}}$ (see the Appendix). Effective potential terms of the dimension-six in the decomposition are 
\begin{eqnarray}
U_{\rm eff}^{(6)}&=&\frac{3}{32 M_S^2 \pi^2} 
(
\frac{1}{3} {\rm tr}({\cal M}_\Lambda^2)^3
-\frac{1}{2 M_S^2} {\rm tr}[({\cal M}_\Gamma^2)^2 ({\cal M}_\Lambda^2)^2] \\
&+& \frac{1}{6 M_S^4} {\rm tr}[({\cal M}_\Gamma^2)^4 {\cal M}_\Lambda^2]
-\frac{1}{60 M_S^6} {\rm tr}({\cal M}_\Gamma^2)^6 
). \nonumber
\end{eqnarray}
Given the Lagrangian of the Higgs boson--squarks interaction (see the Appendix), squark mass matrices $\cal M$ can be calculated and $\kappa_i$ factors in front of the dimension-six terms can be derived. For example, factors $\kappa_1$ and $\kappa_2$ written in the form which uses powers of ($\mu/M_S$) and ($A/M_S$) are 
\begin{eqnarray}
\kappa_1 &=& \frac{h_D^6}{32 M_S^2 \pi^2} \left(2-\frac{3 |A_D|^2}{M_S^2}+\frac{|A_D|^4}{M_S^4}-\frac{|A_D|^6}{10 M_S^6} \right) \\
&-& h_D^4 \frac{g_1^2+g_2^2}{128 M_S^2 \pi^2} \left( 3-3\frac{|A_D|^2}{M_S^2}+\frac{|A_D|^4}{2 M_S^4} \right) \nonumber\\
&+& \frac{h_D^2}{512 M_S^2 \pi^2} \left(\frac{5}{3}g_1^4+2 g_1^2 g_2^2+3g_2^4 \right) \left(1-\frac{|A_D|^2}{2 M_S^2} \right) \nonumber\\
&-&h_U^6 \frac{|\mu|^6}{320 M_S^8 \pi^2}
+h_U^4 \frac{(g_1^2+g_2^2)|\mu|^4}{256 M_S^6 \pi^2}
-h_U^2 \frac{(17 g_1^4-6g_1^2 g_2^2+9g_2^4) |\mu|^2}{3072 M_S^4 \pi^2} \nonumber \\
&+&\frac{g_1^2}{1024 M_S^2 \pi^2} (g_1^4-g_2^4), \nonumber
\end{eqnarray}
\begin{eqnarray}
\kappa_2 &=& -h_D^6 \frac{|\mu|^6}{320 M_S^8 \pi^2} 
+ h_D^4 \frac{(g_1^2+g_2^2)|\mu|^4}{256 M_S^6 \pi^2} 
- h_D^2 \frac{(5 g_1^4+6g_1^2 g_2^2+9g_2^4) |\mu|^2}{3072 M_S^4 \pi^2} \\
&-&\frac{h_U^6}{32 M_S^2 \pi^2} \left(-2+\frac{3 |A_U|^2}{M_S^2}-\frac{|A_U|^4}{M_S^4}+\frac{|A_U|^6}{10 M_S^6} \right)\nonumber \\
&-& h_U^4 \frac{g_1^2+g_2^2}{128 M_S^2 \pi^2} \left( 3-3\frac{|A_U|^2}{M_S^2}+\frac{|A_U|^4}{2 M_S^4} \right) \nonumber \\
&+&\frac{h_U^2}{3072 M_S^2 \pi^2} \left(17 g_1^4-6 g_1^2 g_2^2+9g_2^4 \right) \left(2-\frac{|A_U|^2}{M_S^2} \right) \nonumber \\
&-&\frac{g_1^2}{1024 M_S^2 \pi^2} (g_1^4-g_2^4).  \nonumber
\end{eqnarray}
In a more compact notation $\kappa_i$, $i=$1,...13 can be rewritten using gauge coupling dependent factors $G_i$, $i=$1,...4, and parameter dependent factors $A_j$, $B_k$ and $C_l$ 
\begin{eqnarray}
\kappa_1 &=& h_D^6 C_9^D-h_D^4 G_4 C_8^D+h_D^2 G_2 B_1^D+h_U^6 A_1+h_U^4 G_4 A_2-h_U^2 G_3 A_3+G_1,\\
\kappa_2 &=& h_D^6 A_1+h_D^4 G_4 A_2-h_D^2 G_2 A_3+h_U^6 C_9^U-h_U^4 G_4 C_8^U+h_U^2 G_3 B_1^U-G_1,\\
\kappa_3 &=& h_D^6 C_7^D+h_D^4 G_4 B_3^D-h_D^2 G_2 (2B_1^D+A_3)\\
&+& h_U^6 C_1^U-h_U^4 G_4 B_4^U |\mu |^2+h_U^2 G_3 (B_1^U+2A_3)- 3G_1,\nonumber \\ 
\kappa_4 &=& h_D^6 C_1^D-h_D^4 G_4 B_4^D |\mu |^2+h_D^2 G_2 (B_1^D+2A_3)\\
&+& h_U^6 C_7^U+h_U^4 G_4 B_3^U-h_U^2 G_3 (2 B_1^U+A_3)+3G_1,\nonumber \\
\kappa_5 &=& h_D^6 C_7^D+h_D^4 G_4 B_3^D-h_D^2 G_2 (2 B_1^D+A_3)\\
&+& h_U^6 C_1^U-h_U^4 G_4 B_4^U |\mu |^2+h_U^2 G_3 (B_1^U+2A_3)-3G_1, \nonumber\\
\kappa_6 &=& h_D^6 C_1^D-h_D^4 G_4 B_4^D |\mu |^2+h_D^2 G_2 (B_1^D+2 A_3)\\
&+&h_U^6 C_7^U+h_U^4 G_4 B_3^U-h_U^2 G_3 (2 B_1^U+A_3)+3 G_1, \nonumber\\
\kappa_7 &=& \frac{\mu^3}{320 M_S^8 \pi^2} (A_D^3 h_D^6+A_U^3 h_U^6), \\
\kappa_8 &=& h_D^6 C_6^D+2 h_D^4 G_4 C_4^D+h_D^2 G_2 A_7^D+h_U^6 A_2^U+h_U^4 G_4 A_5^U+h_U^2 G_3 A_7^U,\\
\kappa_9 &=& h_D^6 C_2^D-h_D^4 G_4 A_6^D+h_U^6 A_4^U+h_U^4 G_4 A_6^U,\\
\kappa_{10} &=& h_D^6 A_4^D+h_D^4 G_4 A_6^D+h_U^6 C_2^U-h_U^4 G_4 A_6^U,\\ 
\kappa_{11} &=& h_D^6 C_3^D+h_D^4 G_4 C_5^D-2 h_D^2 G_2 A_7^D+h_U^6 C_3^U+h_U^4 G_4 C_5^U-2 h_U^2 G_3 A_7^U,\\
\kappa_{12} &=& h_D^6 A_2^D+h_D^4 G_4 A_5^D+h_D^2 G_2 A_7^D+h_U^6 C_6^U+2 h_U^4 G_4 C_4^U+h_U^2 G_3 A_7^U,\\
\kappa_{13} &=& h_D^6 C_3^D+h_D^4 G_4 C_5^D-2 h_D^2 G_2 A_7^D+h_U^6 C_3^U+h_U^4 G_4 C_5^U-2h_U^2 G_3 A_7^U,
\end{eqnarray}
where ($X=U,D$)
\begin{equation}
G_1=\frac{1}{M_S^2}\frac{g_1^2(g_1^4-g_2^4)}{1024 \pi^2}, \qquad
G_2=\frac{5g_1^4+6g_1^2g_2^2+9g_2^4}{3072 \pi^2},
\end{equation}
\[ G_3=\frac{17 g_1^4-6g_1^2 g_2^2+9g_2^4}{3072 \pi^2},\qquad G_4=\frac{g_1^2+g_2^2}{256 \pi^2},\]
\begin{equation}
A_1=-\frac{|\mu |^6}{320 M_S^8 \pi^2}, \qquad
A_2=\frac{|\mu |^4}{M_S^6}, \qquad A_3=\frac{|\mu |^2}{M_S^4}, \qquad
A_2^X=\frac{3 A_X \mu |\mu|^4}{320 M_S^8 \pi^2},
\end{equation}
\[A_4^X=-\frac{3 A_X^2 \mu^2 |\mu|^2}{320 M_S^8 \pi^2}, \qquad A_5^X=-\frac{2 A_X \mu |\mu|^2}{M_S^6}, \qquad
A_6^X=\frac{A_X^2 \mu^2}{M_S^6}, \qquad
A_7^X=\frac{\mu A_X}{M_S^4}, \]
\begin{equation}
B_1^X=-\frac{|A_X|^2}{M_S^4}+\frac{2}{M_S^2}, \qquad
B_2^X=-\frac{4 |A_X|^2}{M_S^6}+\frac{6}{M_S^4}, 
\end{equation}
\[B_3^X=C_8^X+|\mu|^2 B_2^X, \qquad
B_4^X=\frac{|\mu|^2}{M_S^6}+B_2^X, \]
\begin{equation}
C_1^X=\frac{|\mu|^4}{320 \pi^2} \left(-\frac{9|A_X|^2}{M_S^8}+\frac{10}{M_S^6} \right), \qquad
C_2^X=\frac{A_X^2 \mu^2}{320 \pi^2} \left(-\frac{3|A_X|^2}{M_S^8}+\frac{10}{M_S^6}\right),
\end{equation}
\[ C_3^X=\frac{A_X \mu |\mu|^2}{320 \pi^2} \left(\frac{9 |A_X|^2}{M_S^8}-\frac{20}{M_S^6} \right), \qquad
C_4^X=A_X \mu \left(\frac{|A_X|^2}{M_S^6}-\frac{3}{M_S^4} \right),\]
\[C_5^X=-2 A_X \mu \left(\frac{|A_X|^2-|\mu|^2}{M_S^6}-\frac{3}{M_S^4} \right), \qquad
C_6^X=\frac{A_X \mu}{320 \pi^2} \left(\frac{3|A_X|^4}{M_S^8}-\frac{20 |A_X|^2}{M_S^6}+\frac{30}{M_S^4} \right),
\]
\[C_7^X=-\frac{|\mu|^2}{320 \pi^2} \left(\frac{9 |A_X|^4}{M_S^8}-\frac{40|A_X|^2}{M_S^6}+\frac{30}{M_S^4} \right),\qquad
C_8^X=\frac{|A_X|^4}{M_S^6}-\frac{6 |A_X|^2}{M_S^4}+\frac{6}{M_S^2},
\]
\[C_9^X=-\frac{1}{320 \pi^2} \left(\frac{|A_X|^6}{M_S^8}-\frac{10 |A_X^4|}{M_S^6}+\frac{30 |A_X|^2}{M_S^4}-\frac{20}{M_S^2} \right).
\]
Meaningful numerical results following from the effective potential expansions in the inverse powers of $M_S$ are using the assumption of small mass splitting among the squark mass eigenstates (or simultaneous decoupling of squark fields). In the literature it is usually considered that the expansion is valid if
$(m^2_{\tilde{t}_1}-m^2_{\tilde{t}_2})/(m^2_{\tilde{t}_1}+m^2_{\tilde{t}_2}) < 0.5$ where $m_{\tilde{t}_{1,2}}$ are the stop masses. Then $M^2_{S}$ can be defined as the average $(m^2_{\tilde{t}_1}+m^2_{\tilde{t}_2})/2$\footnote{Besides the abovementioned approach developed in \cite{mssm_radcor_new}, recent direct comparison of results for the one-loop MSSM amplitudes $ggh$ and $\gamma \gamma h$ obtained by means of the diagrammatic calculation and the covariant derivative expansion (CDE) method \cite{gaillard} for the case of either degenerate or non-degenerate stop mass spectrum, can be found in \cite{drozd}. For large $\tan \beta$ the approximation of (almost) degenerate stop masses is not satisfactory at $m_{\tilde{t}}<$0.5 TeV and large $X_t$ mixing parameter values of a few TeV, however, $m_h=$125 GeV is mostly available.}. The contribution of dimension-six operators is small in the phase with softly broken symmetry if at least $2|m_{top} A_t|<M^2_S$ and $2|m_{top} \mu|<M^2_S$ \cite{carena95}. However, the dimension-six terms may play an important role in the $A,\mu$ parameter range of about/of the order of 10$^1$ TeV and moderate $M_S$. For example, values of $\kappa_i$ evaluated for $A=10$ TeV, $\mu=14$ TeV, $\tan \beta=5$ are shown in Fig. \ref{fig-1}, where the dimensionless couplings $\kappa_i \cdot M_S^2$ are depicted for $M_S$ values of 5 and 7 TeV. The behavior of $\lambda_i$ and $\kappa_i \cdot M_S^2$ as a function of $M_S$ at the multi-TeV energy scale is shown in Fig. \ref{fig-2}. One can see that significant values of $\kappa_i \cdot M_S^2$ are observed in $M_S$ range less than 8 TeV.

\begin{figure*}[h]
\begin{minipage}[h]{0.5\linewidth}
\center{\includegraphics[width=1\linewidth]{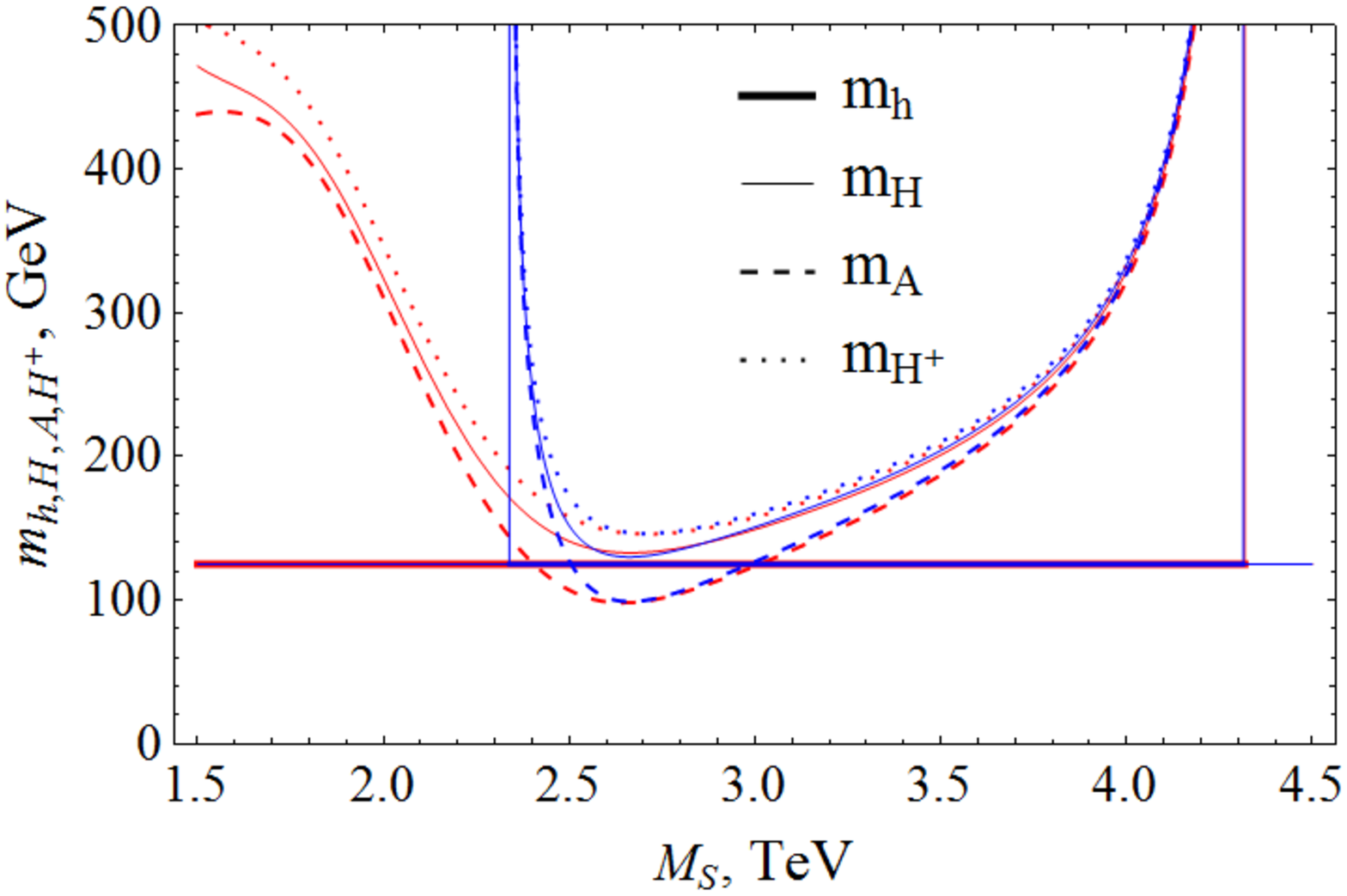}} (a)
\end{minipage} 
\begin{minipage}[h]{0.5\linewidth}
\center{\includegraphics[width=1\linewidth]{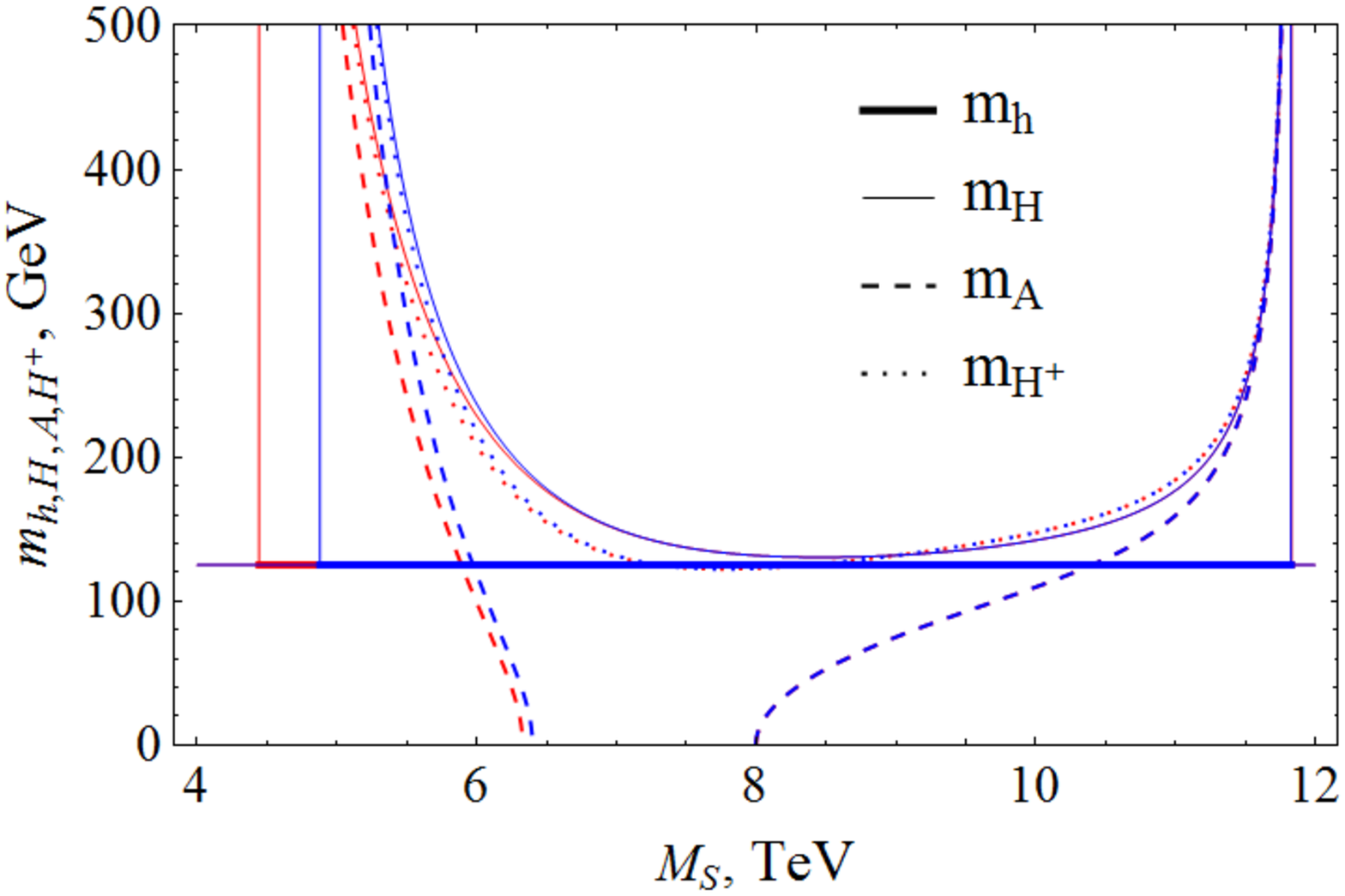}} (b)
\end{minipage} 
\begin{minipage}[h]{0.5\linewidth}
\center{\includegraphics[width=1\linewidth]{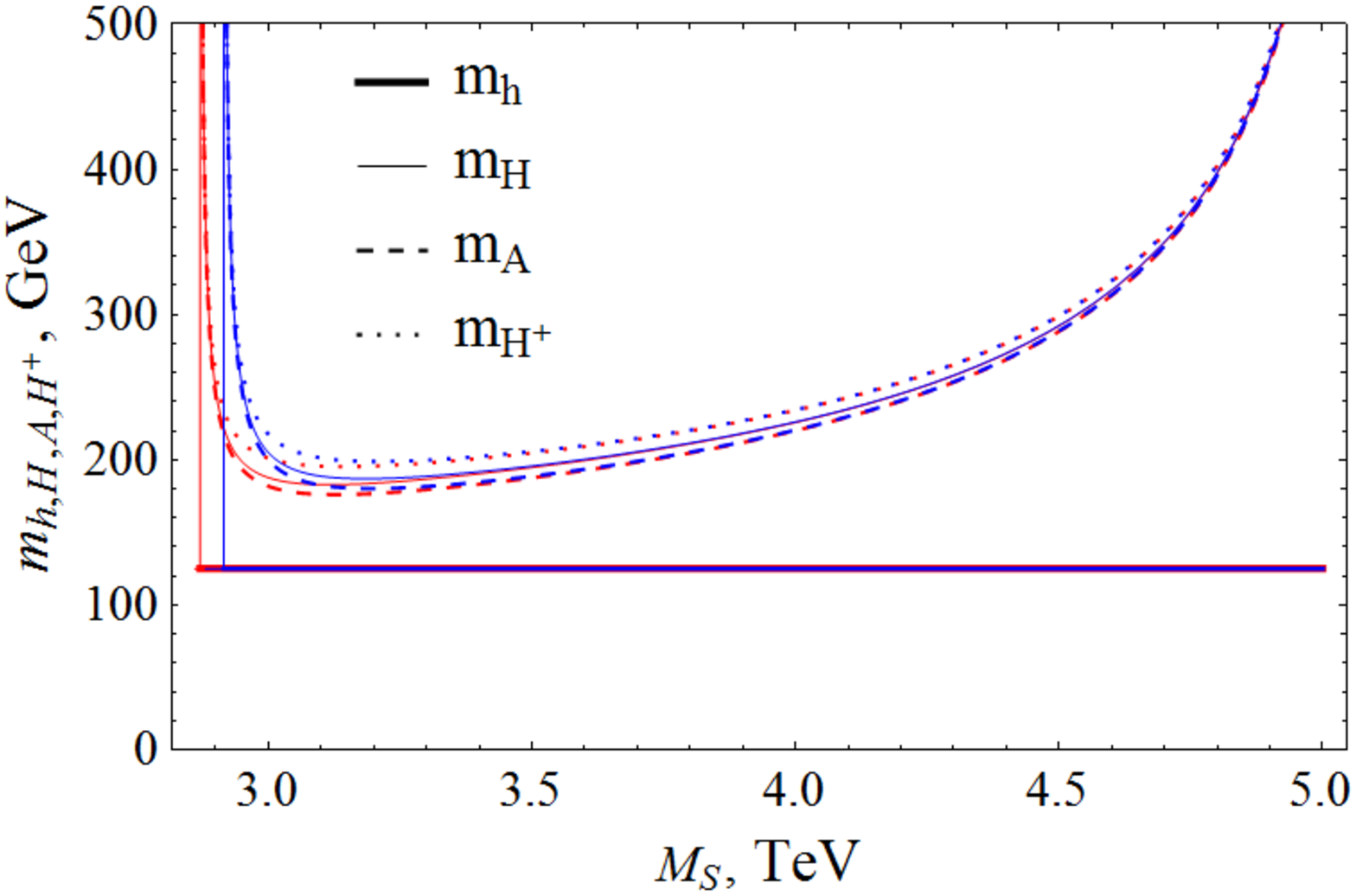}} (c)
\end{minipage} 
\begin{minipage}[h]{0.5\linewidth}
\center{\includegraphics[width=1\linewidth]{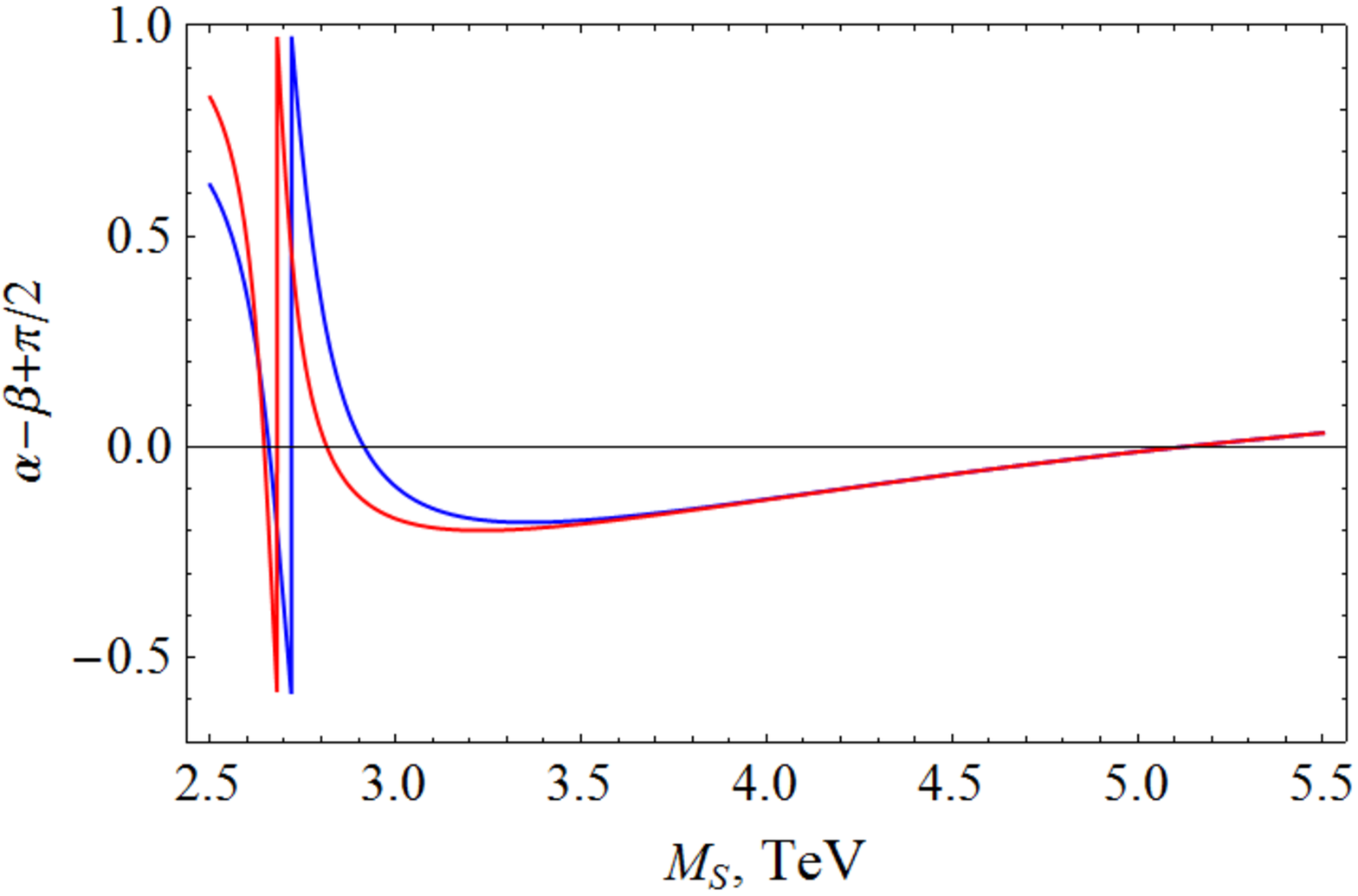}} (d)
\end{minipage} 
\caption{
Higgs boson masses and mixing angles combination as functions of the squark mass scale $M_S$: $m_h$ (thick horizontal line), $m_H$ (thin line), $m_A$ (dashed line) and $m_{H^\pm}$ (dotted line). Red lines correspond to the effective potential $U^{(4)}$ including terms with the maximal dimension four in the fields, blue lines are results for masses calculated with the effective potential $U^{(6)}$ including the dimension-six operators. The MSSM parameter sets: (a) $\tan \beta=4$, $A=10$ TeV, $\mu=8$ TeV; (b) $\tan \beta=8$, $A=25$ TeV, $\mu=30$ TeV, (c),(d) $\tan \beta=$5, $A=$10 TeV and $\mu=$5 TeV. The discontinuity in Fig. (c) at $M_S$ of about 3 TeV corresponds to zero denominator of Eq.(\ref{m2A}).  
}

\label{fig-3} 
\end{figure*}

\begin{figure*}[h]
\begin{minipage}[h]{0.45\linewidth}
\center{\includegraphics[width=1\linewidth]{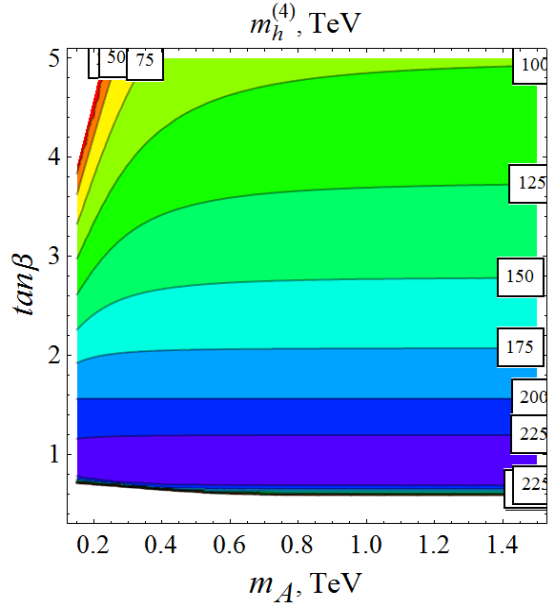}} (a)
\end{minipage} 
\hfill
\begin{minipage}[h]{0.45\linewidth}
\center{\includegraphics[width=1\linewidth]{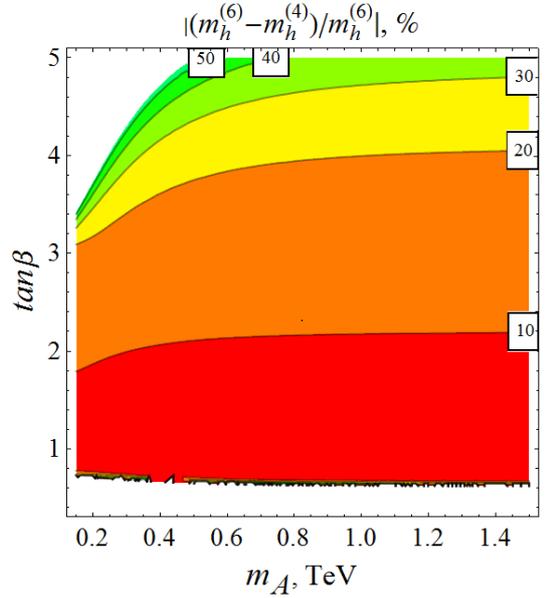}} (b)
\end{minipage} 
\hfill
\begin{minipage}[h]{0.45\linewidth}
\center{\includegraphics[width=1\linewidth]{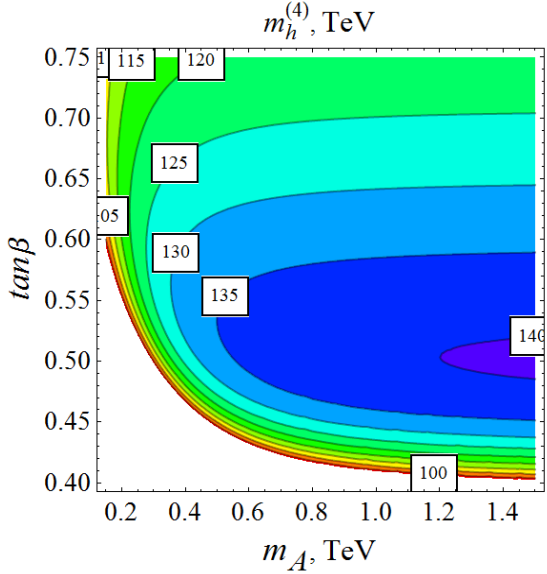}} (c)
\end{minipage}
\hfill 
\begin{minipage}[h]{0.45\linewidth}
\center{\includegraphics[width=1\linewidth]{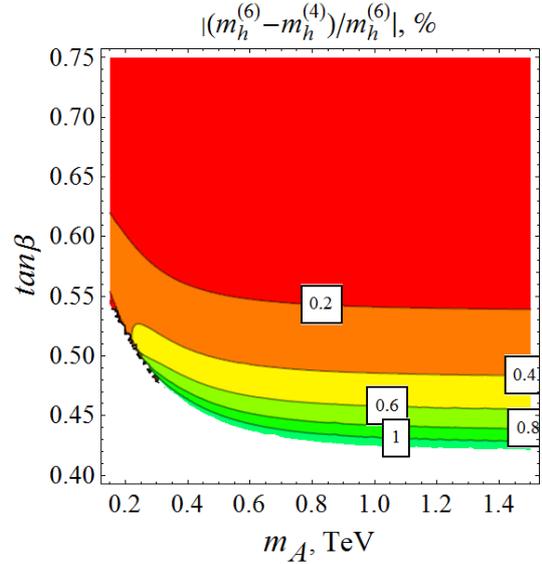}} (d)
\end{minipage} 
\caption{
Left panel: contours for the Higgs boson mass $m_h^{(4)}$ calculated with the dimension-four potential terms; right panel: the relative difference in percent between $m_h^{(6)}$ and $m_h^{(4)}$ masses; the parameter set $A=10$ TeV, $\mu=8.3$ TeV, $M_S=2$ TeV (a, b) and $M_S=5$ TeV (c, d).
}
\label{fig-4} 
\end{figure*}

\begin{figure*}[h]
\begin{center}
\begin{minipage}[h]{0.4\linewidth}
\center{\includegraphics[width=1\linewidth]{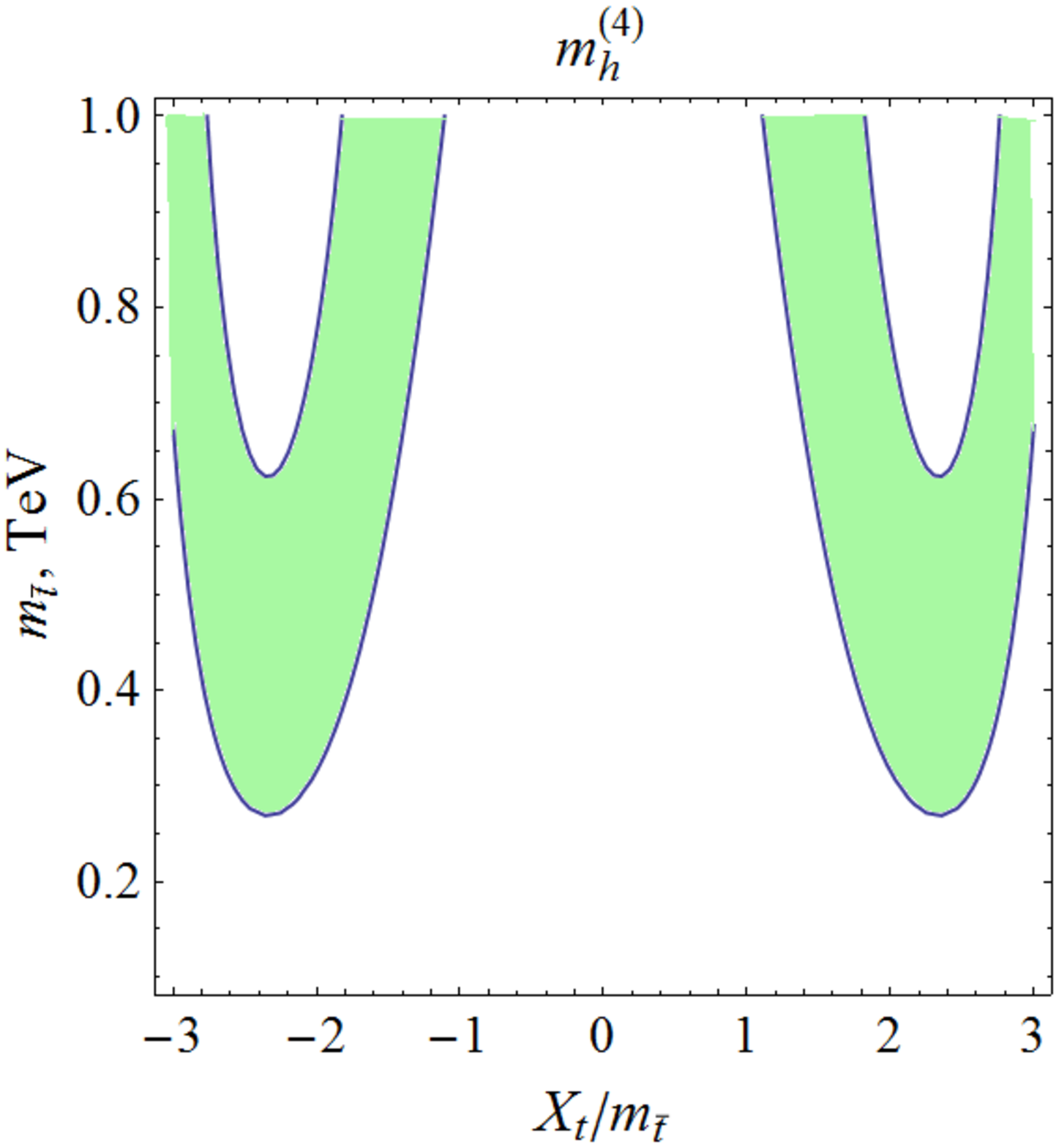}} (a)
\end{minipage} 
\begin{minipage}[h]{0.4\linewidth}
\center{\includegraphics[width=1\linewidth]{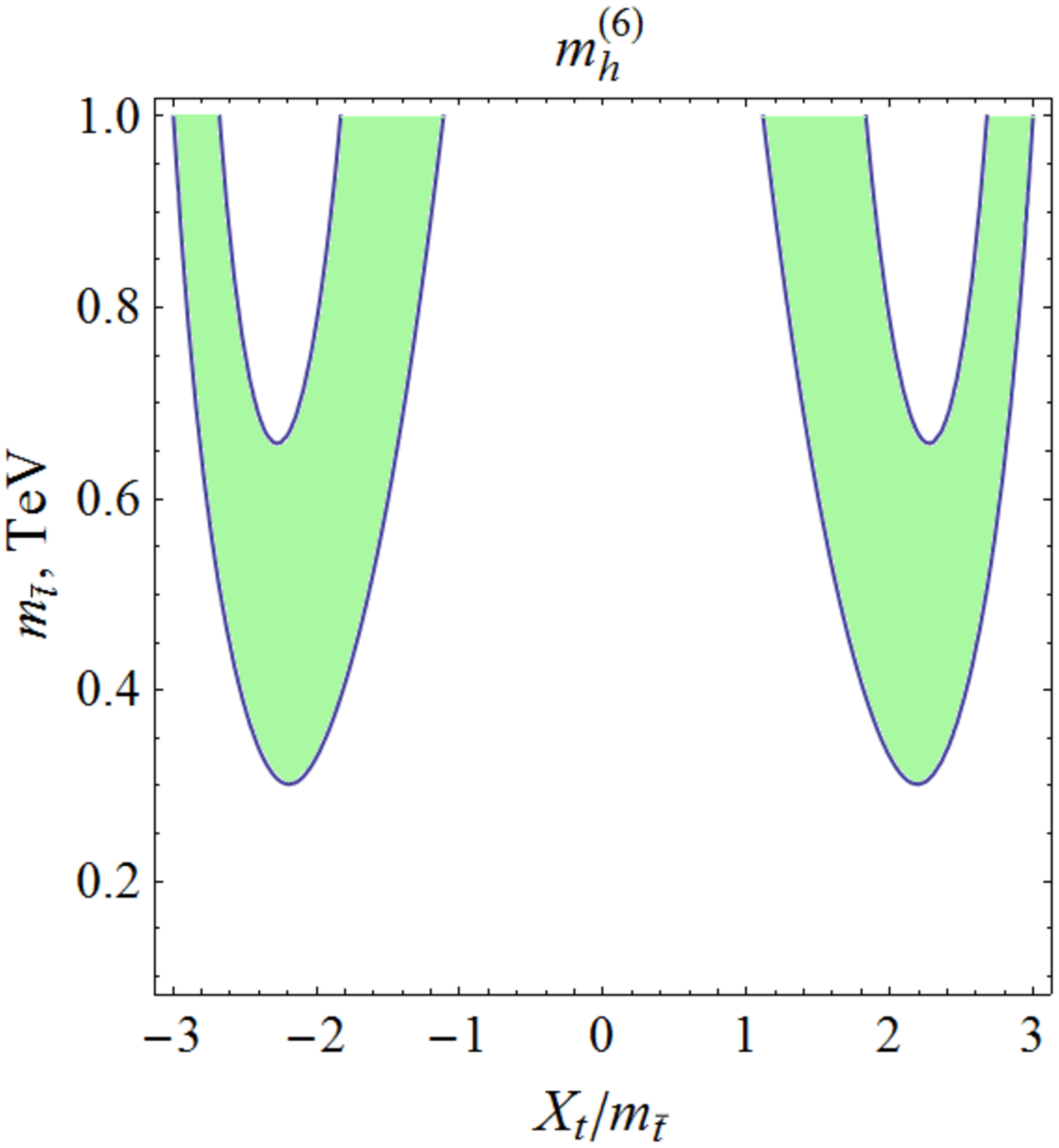}} (b)
\end{minipage}
\end{center}
\caption{
Domains of the Higgs boson mass $m_h=$125$\pm$3 GeV for $m_A=300$ GeV and the Higgs superfield mass parameter $\mu$ equal to zero, $\tan \beta=20$,
calculated with the dimension-four operators (a) and the dimension-six operators (b).
}
\label{fig-5} 
\end{figure*}

The Higgs boson masses $m_{H,A,H^\pm}$ evaluated for two ($\tan \beta$, $A$, $\mu$) parameter sets at fixed value of the lightest $CP$-even state mass $m_h=125$ GeV and large $X_t$ mixing parameter of the order of 10 TeV are shown in Fig. \ref{fig-3} as a function of the squark mass scale $M_S$.
The $CP$-odd scalar mass $m_A$ is calculated using Eq.(\ref{m2A}), where $m_h$ is an input parameter with fixed value. A pole of $m^2_A(M_S)$ may take place when the denominator in Eq.(\ref{m2A}) is zero. 
In the unphysical region of $M_S$, for example, to the left from the pole in Fig. \ref{fig-3}(c), the restrictions imposed by Eq.(\ref{2cond}) are not respected.
Contribution of the dimension-six terms $U^{(6)}$ to masses of scalars is very small in comparison with the dimension-four terms $U^{(4)}$ for moderate $M_S$ ($M_S \geq 3$ TeV, Fig. \ref{fig-3}(a) and $M_S \geq 7$ TeV, Fig. \ref{fig-3}(b)) but for smaller $M_S$ corrections are very important. In Fig. \ref{fig-3}(a) the physical region of $m^2_h>$0 indicated by vertical lines narrows to 2.3 TeV (lower bound). 
In the case (b), see Fig. \ref{fig-3}, the $CP$-odd scalar mass squared is not positively defined for $M_S$ range from 6.3 to 8 TeV. At moderate $\tan \beta \approx $10 positively defined masses squared of $H$, $A$ and $H^\pm$ consistent with the input $m_h=$125 GeV are not possible for $M_S$ greater than 12 TeV. Note that nonstandard mass spectrum with extremely light pseudoscalar is available in this case. At fixed $m_h=$125 GeV the $CP$-odd state $A$ can be as light as 25--30 GeV with $H$ and $H^\pm$ states in the decoupling regime or with masses of the order of electroweak scale. For example, Higgs masses for (b) parameter set, see Fig. \ref{fig-3}, and $M_S\simeq 6.3$ TeV are $m_h=125$ GeV, $m_H=190$ GeV, $m_A=27$ GeV, $m_{H^\pm}=170$ GeV. The alignment limit when $\alpha \approx \beta-\pi/2$ takes place for set (b) in the vicinity of $M_S=$5.5 TeV; it is possible for $A, H, H^\pm$ in the decoupling regime only. The regime of alignment without decoupling without small $m_A$ is available if $\tan \beta=$5, $A=$10 TeV and $\mu=$5 TeV.
For this parameter set, see Fig. \ref{fig-3}(c), when curves are more stable with respect to corrections, there are two alignment limits. In Fig. \ref{fig-3}(d) the first alignment limit takes place at $M_S=$2.98 TeV without decoupling and the second limit at 5.1 TeV demonstrates decoupling of $H$,$A$ and $H^\pm$ states.   
Figure \ref{fig-4} illustrates an increasing role of corrections from the $U^{(6)}$ terms to $m_h$ in the case of 'low $\tan \beta$' scenarios \cite{limited} which are found to be of about 1\% at $M_S=5$ TeV and $A, \mu$ more than 10 TeV and of around 20\% for the lower superpartner mass scale $M_S=2$ TeV and $A, \mu$ less than 10 TeV. In Fig. \ref{fig-5} the condition $m_h=$125$\pm$3 GeV is translated to the mixing parameter -- quark superpartner plane $(X_t/M_S, M_S)$, demonstrating sensitivity of the contours in the regime $\mu=$0 (see also \cite{drozd}, where similar contours are reconstructed using the diagrammatic calculation \cite{feynhiggs}). Increasing $\mu$ parameter of a few hundreds of GeV changes strongly these exclusion contours, leaving only small acceptable domain in the left upper corner of the plot.

\section*{Summary}

In the absence of direct evidence motivating extensions of the Standard Model-like Higgs sector, the effective field theory approach is a convenient framework to describe possible new physics either in a model-independent or in a model-dependent way. In both cases the MSSM Lagrangian is extended by higher-dimensional operators which are suppressed by the mass scale of new physics. In the model-dependent case of the MSSM when the resummed effective potential is expanded
up to dimension-six operators induced by the soft supersymmetry breaking terms, we calculated symbolically corrections to the effective sextic 
couplings and used them to determine the post-Higgs discovery mass spectrum of the heavy MSSM Higgs bosons. An improved precision can be reached using such procedure especially at the low EFT cut-off scale. 
Corrections to the mass spectrum depend strongly on the domain in the MSSM parameter space and are defined mainly by the quark superpartner mass scale and mixing in the sector of soft SUSY-breaking terms. Even at the moderate mixing parameter values significant contributions to the heavy scalar mass spectrum of the order of 10-20\% induced by the dimension-six operators are found at the squark mass scale $M_S \sim$ 2--3 TeV. Thus, for moderately heavy supersymmetry additional corrections induced by higher order terms in the expansion of the effective potential should be taken into account.
One can observe that in a number of cases the restrictions on the MSSM parameter space are not so much a consequence of the condition $m_h=$125 GeV as the presence of mass basis for the five Higgs bosons, where mass hierarchy is acceptable from experimental point of view and there are no tachyonic states.      

\section*{Acknowledgements}
M.D. is grateful to E. Bagnashi and G. Weiglein for useful discussions.
This work was partially supported by Grant No. 4.W02.16.7989-NSh.

\appendix

\section*{Appendix}

Most general scalar potential, including Higgs boson and one generation of squarks, can be written as \cite{haberhempfling, gunion}
\begin{equation}
{\cal V}^0={\cal V}_M+{\cal V}_\Gamma+{\cal V}_\Lambda+{\cal V}_{\tilde{Q}}, \tag{A1}
\end{equation}
where ${\cal V}_M$ contains mass squark terms, ${\cal V}_\Gamma$ -- $F$-terms, ${\cal V}_\Lambda$ -- $D$-terms of Higgs-squark interactions and ${\cal V}_{\tilde{Q}}$ -- quartic squark interaction terms
\begin{align}
&{\cal V}_M = -\mu_{ij}^2 \Phi_i^\dagger \Phi_j+M_{\tilde{Q}}^2(\tilde{Q}^\dagger \tilde{Q})+M_{\tilde{U}}^2(\tilde{U}^* \tilde{U})+M_{\tilde{D}}^2(\tilde{D}^* \tilde{D}), \tag{A2} \\
&{\cal V}_\Gamma = \Gamma_i^D(\Phi_i^\dagger \tilde{Q}) \tilde{D}+\Gamma_i^U(i \Phi_i^T \sigma_2 \tilde{Q})\tilde{U}+h.c., \tag{A3}\\
&{\cal V}_\Lambda = \Lambda_{ik}^{jl} (\Phi_i^\dagger \Phi_j)(\Phi_k^\dagger \Phi_l)+(\Phi_i^\dagger \Phi_j) [\Lambda_{ij}^Q (\tilde{Q}^\dagger \tilde{Q})+\Lambda_{ij}^U (\tilde{U}^* \tilde{U})+\Lambda_{ij}^D (\tilde{D}^* \tilde{D})] \tag{A4}\\
&+ \overline{\Lambda}_{ij}^Q(\Phi_i^\dagger \tilde{Q})(\tilde{Q}^\dagger \Phi_j)+\frac{1}{2} [\Lambda \epsilon_{ij}(i \Phi_i^T \sigma_2 \Phi_j)\tilde{D}^* \tilde{U} \nonumber
+h.c.]
\end{align}
and $\Gamma, \Lambda$ are determined by the tree-level SUSY relations
\begin{align}
& \Lambda^Q = {\rm diag} [ \frac{1}{4} (g_2^2-g_1^2 Y_Q), h_U^2-\frac{1}{4} (g_2^2-g_1^2 Y_Q)], \tag{A5}\\
& \overline{\Lambda}^Q = {\rm diag} (h_D^2-\frac{1}{2} g_2^2, \frac{1}{2} g_2^2-h_U^2),\tag{A6}\\
& \Lambda^U = {\rm diag}(-\frac{1}{4} g_1^2 Y_U, h_U^2+\frac{1}{4} g_1^2 Y_U),\tag{A7}\\
& \Lambda^D = {\rm diag}(h_D^2-\frac{1}{4} g_1^2 Y_D, \frac{1}{4}g_1^2 Y_D),\tag{A8}\\
& \Lambda = -h_U h_D,\tag{A9}\\
& \Gamma^{U}_{1,2}=h_U(-\mu, A_U), \qquad
\Gamma^D_{1,2}=h_D(A_D, -\mu), \tag{A10}
\end{align}
$g_{1,2}$ are couplings of $SU(2)_{L} \times U(1)_Y$, $Y_{Q,U,D}=\{ \frac{1}{3} (-1), \frac{2}{3} (2), -\frac{4}{3} \}$ -- squark (slepton) hypercharges, $h_{U}=\frac{g_2 m_{U}}{\sqrt{2}m_W s_\beta}, h_{D}=\frac{g_2 m_{D}}{\sqrt{2}m_W c_\beta}$ -- Yukawa couplings, $A_{U,D}$ -- trilinear couplings, $\mu$ -- Higgs superfield mass parameter.

The squark mass matrix is obtained by taking derivatives
\begin{equation}
{\cal M}^2_{a,b}=\frac{\partial^2 {\cal V}^0}{\partial \Psi_a \partial \Psi_b^*}, \tag{A11}
\end{equation}
where $\Psi = (\tilde{Q}, \tilde{U}^*, \tilde{D}^*)$.


\end{document}